\documentclass[aps,prb,floatfix,12pt]{revtex4}
\usepackage{graphicx,graphics}\textheight=25cm\topmargin-1.7cm
\begin{document}
\title{On the temperature dependence of ballistic Coulomb drag in nanowires }
\author{M. I. Muradov and V. L. Gurevich }
\affiliation{\it A. F. Ioffe Institute of Russian Academy of
Sciences, 194021 Saint Petersburg, Russia}

\begin{abstract}
We have investigated within the theory of Fermi liquid
dependence of Coulomb drag current in a passive quantum wire on
the applied voltage $V$ across an active wire and on the temperature $T$ for any values of $eV/k_BT$.
We assume that the bottoms of the 1D minibands in both wires
almost coincide with the Fermi level. We come to conclusions that 1)
within a certain temperature interval the drag current can be a
descending function of the temperature $T$; 2) the experimentally
observed temperature dependence $T^{-0.77}$ of the drag current
can be interpreted within the framework of Fermi liquid theory; 3)
at relatively high applied voltages the drag current as a function
of the applied voltage saturates; 4) the screening of the electron
potential by metallic gate electrodes can be of importance.
\end{abstract}
\maketitle

Coulomb drag predicted by Pogrebinskii~\cite{POG} (see also
Price~\cite{Price}) is a phenomenon directly associated with
Coulomb interaction of the electrons in a semiconductor.
Perpetually advancing progress in semiconductor lithography
technique has provided extensive investigation of this effect
between 2D electron layers separated by hundreds or tens of
angstroms and supported interest to this field (see,
e.g.~\cite{Rojo}, where a number of papers dealing with the
Coulomb drag between two 2D layers is discussed).

Coulomb drag effect for two parallel quantum wires in the
ballistic regime has been investigated by Gurevich, Pevzner and
Fenton in Ref.~\onlinecite{GPF} for $eV\,\ll\,k_BT$. This case
may be called linear as the drag current is a linear function of
the applied voltage $V$. The authors of the present paper treated in Ref.~\onlinecite{GM_Nonohmic_Coul} a nonlinear case where $eV\,\gtrsim\,k_BT$. In both cases the Fermi energy
$\mu$ was assumed to be much larger than $k_BT$. As is well known, under this condition the transport phenomena are determined by a stripe of the width $k_BT$ near the Fermi level. This means that the drag current $J_{\rm drag}$ should have a maximum provided the positions of these stripes in both quantum wires coincide. For identical wires (the case treated in Refs.~\cite{GPF,GM_Nonohmic_Coul}) this requirement means coincidence of the bottoms of 1D bands of transverse quantization. Under these conditions $J_{\rm drag}$ goes up with
temperature.

One encounters an entirely different situation provided the Fermi level is near the bottom of a 1D band. If this is the case for two bands in both wires all the electrons of these bands can take part in the electron-electron collisions. The very effect of drag is strongly dependent on the transferred (quasi)momentum $p_{\rm tr}$ in the course of interwire Coulomb scattering of electrons. The number of electrons involved goes down with $p_{\rm tr}$ whereas the interaction responsible for the drag goes up and its influence is predominant for $J_{\rm drag}$. Our purpose is to investigate this situation. In other words, we assume that
\begin{equation}
\vert\mu_n\vert\lesssim k_BT,\quad \mu_n\equiv\mu-\varepsilon_n(0)
\end{equation}
where $\varepsilon_n(0)$ is the position of the bottom of $n$th 1D band (a result of the transverse quantization) and $\mu$ is the chemical potential. In the discussion of the
experimental situation we will use the findings of
Refs.~\cite{DVRPRM} and~\cite{DZRKVN} (see also the review
paper~\cite{DGKN}). As is seen in these works, the drag voltage
peaks occur just where the quantized conductance of the drive
(active) wire  rises between the plateaus, i.e. the maximum of the drag effect occurs provided the 1D bands of the two quantum
wires are aligned (i. e. their bottoms coincide within the
accuracy of $\lesssim k_BT$) and Fermi quasimomenta are small.
First two peaks are well pronounced. First of them corresponds to
alignment of the two ground 1D bands in both wires. The second one
corresponds to alignment of the ground (first) 1D band of the
passive (drag) wire and the second 1D band of the drive wire.

It was found that the temperature dependence of the drag current
can be described by the law $\sim\,T^{-0.77}$ in the temperature
interval from $60$\,mK to $1$ K. The authors of these papers
highlight this temperature dependence claiming that the power-law
temperature dependence of the drag resistance is a signature of
the Luttinger liquid state. The authors of~\cite{PBT} evidently
share the same opinion, claiming that for coupled Fermi liquid
systems the drag resistance always is an increasing function of
temperature.

In this paper we will argue that the situation is not so simple,
and the temperature dependence observed on experiment can be
explained within the Fermi liquid approach.

Experimentally found magnitudes of the drag resistance are of
order of hundreds Ohms or even smaller and one should provide a
special explanation for the weakness of the interwire electron-electron interaction.
We believe that it is due to
the screening of Coulomb interaction by the gates.
Such screening has not been taken into consideration so far.
In Ref.~~\cite{GM_Nonohmic_Coul} the following equation has been derived for the drag current
(see Eq. (13) in~\cite{GM_Nonohmic_Coul})
\begin{equation}
\label{nobddrag_cur} J_{\rm
drag}=-2e\sinh{\left({eV\over{2k_BT}}\right)} {2\pi\over{\hbar}}
{mL\over{2\pi\hbar}} \left({2L\over{2\pi\hbar}}\right)^2
\left({2e^2\over {\kappa L}}\right)^2
\sum_{nn^{\prime}}\int_0^{\infty}dp\int_0^{\infty}dp^{\prime}
\displaystyle{g_{nn^{\prime}}(p+p^{\prime})\over{p+p^{\prime}}}
{\cal Q},
\end{equation}
where
\begin{equation}
\label{nobd4} {\cal Q}=\exp{\varepsilon^{(1)}_{np}-\mu\over k_BT}
\exp{\varepsilon^{(2)}_{n\prime p\prime}-\mu\over k_BT}
f(\varepsilon^{(1)}_{np}-\mu)f\left(\varepsilon^{(2)}_{n^{\prime}p^{\prime}}-
\mu-{eV\over2}\right)f(\varepsilon^{(1)}_{np^{\prime}}-\mu)
f\left(\varepsilon^{(2)}_{n^{\prime}p}-\mu+{eV\over2}\right),
\end{equation}
Here $f(\varepsilon-\mu)$ is the Fermi function and $\kappa$ is the dielectric susceptibility.

The unscreened Coulomb interaction matrix element squared
 $g_{nl}(p+p^{\prime})$ can be written as $\left[K_0(d(p+p^{\prime})/\hbar)\right]^2$
provided the widths of the wires are much smaller than the interwire distance $d$.
Here $K_0(s)$ is the  MacDonald function. Using the random phase approximation one
can straightforwardly take into consideration the screening by the gates as well as
by the quantum wires themselves. The resulting equation being, however, too cumbersome,
we will take into account the screening only by the gates treating them as a single plane. As for
the contribution of 1D wires to the screening, we will neglect it. As a result,
we get for the screened Coulomb interaction
\begin{eqnarray}\label{dv14}
U_s(\omega,q_x)=\int\frac{d{\bf q}_{\perp}}{(2\pi)^2}C_n({\bf
q}_{\perp})U_{\bf q}C_l(-{\bf q}_{\perp})+\nonumber\\
+\int\frac{d{\bf q}_{\perp}}{(2\pi)^2}C_n({\bf q}_{\perp})U_{\bf
q}\int\frac{dq_z^{\prime}}{2\pi}C_l(-q_y,-q_z^{\prime})
U_{q_x,q_y,q_z^{\prime}}\frac{\Pi^R_{\omega}({q_x,q_y})}
{1-\Pi^R_{\omega}({q_x,q_y})U(q_x,q_y)}
\end{eqnarray}
where $C_n({\bf q}_{\perp})=\langle n|e^{i{\bf q}_{\perp}{\bf
r}_{1\perp}}|n\rangle$ and $C_l({\bf q}_{\perp})=\langle l|e^{i{\bf q}_{\perp}{\bf r}_{2\perp}}|l\rangle$. Here $|n\rangle$ and $|l\rangle$ are the
transverse wave functions of the first and second quantum wires.
Precisely,
$$
|n\rangle=\phi_{np}({\bf r}_{\perp}),\quad {\bf r}_{\perp}\longleftrightarrow y,z
$$
are the wave function describing the transverse quantization.
$U_{\bf q}=4\pi e^2/\kappa q^2$ is the Fourier transform of the
3D Coulomb potential, $U(q_x,q_y)=\int\,dq_z\,U_{\bf q}/2\pi$. Polarization operator
$\Pi^R(\omega,q_x,q_y)$ for a 2D layer can be found
in Ref.~\cite{Stern}. We assume that the gate electrodes are made
of a metal where the period of plasma vibration is much shorter than
any characteristic time of a semiconductor. Therefore  we will deal
only with a static as well as long wave
limit of this operator. In this limit it is reduced to the
2D electron density of states.

We assume that the gates are in the plane $z=0$, two quantum wires parallel
to the plane (and oriented along the $x$-axis) are displaced by the same
distance  $z_0$, the interwire distance being $d$. For the electrons with
coordinates $x,0,z_0$ and $x^{\prime},d,z_0$, belonging to two wires
\begin{eqnarray}\label{dv16}
U_s(\omega,x-x^{\prime})=e^2\int\frac{e^{-iq_yd-iq_x(x-x^{\prime})}dq_ydq_x}{2\pi\sqrt{q_x^2+q_y^2}}+\nonumber\\
+e^2\int\frac{e^{-iq_yd-iq_x(x-x^{\prime})}dq_ydq_x}{2\pi(q_x^2+q_y^2)}\frac{2\pi
e^2\Pi^R_{\omega}({q_x,q_y})e^{-2\sqrt{q_x^2+q_y^2}z_0}}{1-2\pi
e^2\Pi^R_{\omega}({q_x,q_y})/\sqrt{q_x^2+q_y^2}}.
\end{eqnarray}
In the static case we arrive at a simple result
\begin{eqnarray}\label{dv17}
U_s(x-x^{\prime})=\frac{e^2}{\sqrt{(x-x^{\prime})^2+d^2}}-\frac{e^2}{\sqrt{(x-x^{\prime})^2+d^2+(2z_0)^2}},
\end{eqnarray}
the second term here describes the action of an "image" (we assume that $z_0$ is bigger than the Bohr radius).

Therefore we get for the drag current
instead of (\ref{nobddrag_cur})
\begin{equation}
\label{nobdwithscr} J_{\rm
drag}=-\frac{2e^5mL}{\pi^2\hbar^4\kappa^2}\left(\frac{z_0}{d}\right)^4\left(\frac{d}{\hbar}\right)^2
\sinh{\frac{eV}{2k_BT}}\sum_{nl}\int_0^{\infty}dp\int_0^{\infty}dp^{\prime}(p+p^{\prime})
K_1^2\left(d(p+p^{\prime})/\hbar\right) {\cal Q},
\end{equation}
where now
\begin{eqnarray*}
\label{nobd4withscr}
{\cal Q}=\frac{1}{\cosh{[(p^2-p_n^2)/4mk_BT]}\cosh{[(p^2-p_n^2+meV-2m\varepsilon_{nl})/4mk_BT]}}\\
\times \frac{1}{\cosh{[(p^{\prime
2}-p_n^2)/4mk_BT]}\cosh{[(p^{\prime
2}-p_n^2-meV-2m\varepsilon_{nl})/4mk_BT]}}
\end{eqnarray*}
and
$$
\varepsilon_{nl}=\varepsilon_{n}-\varepsilon_{l}.
$$
Here we have defined the Fermi quasimomentum
$p_n=\sqrt{2m(\mu-\varepsilon_n)}$ while $\varepsilon_{n}$ and $\varepsilon_{l}$
are the positions of 1D band bottoms. In what follows we will assume that
\begin{equation}
d\gg z_0,
\end{equation}
i. e. the spacing $d$ between the quantum wires is larger than their distance $z_0$ to the gates.

Now
\begin{equation}
K_0\left(d\frac{p+p^{\prime}}{\hbar}\right)-K_0\left(\sqrt{1+(2z_0/d)^2}d\frac{p+p^{\prime}}{\hbar}\right)
\simeq\,2\left(\frac{z_0}{d}\right)^2d\frac{p+p^{\prime}}{\hbar}K_1\left(d\frac{p+p^{\prime}}{\hbar}\right).
\end{equation}
The scale of variation of $Q$ as a function of $p$ and $p^{\prime}$
is the thermal momentum $\sqrt{4mk_BT}$. At the same time
$(p+p^{\prime})K_1^2\left(d(p+p^{\prime})/\hbar\right)$ is a rapidly
decreasing function, the scale of its variation is $\hbar/d$. For
$$\hbar/d\,\ll\,\sqrt{4mk_BT}$$ one can take out of the integral all
the slowly varying functions keeping as the integrand only
$(p+p^{\prime})K_1^2\left(d(p+p^{\prime})/\hbar\right)$. For
$p_n\geq\,\hbar/d$ in the case of 1D band alignment in two wires
$$\varepsilon_{n}\approx\varepsilon_{l}$$ we can retain the contribution
only of these 1D bands in the sum and get
\begin{equation}
\label{nobdwithscr1} J_{\rm drag}=J_0
\frac{\sinh{[eV/2k_BT]}}{\cosh^2{[p_n^2/4mk_BT]}\cosh{[(p_n^2-meV)/4mk_BT]}\cosh{[(p_n^2+meV)/4mk_BT]}},
\end{equation}
where
\begin{equation}
J_0=-\frac{3e^5mL}{16\hbar^3d\kappa^2}\left(\frac{z_0}{d}\right)^4.
\end{equation}
Here we have made use of the equation
\begin{equation}
\int_0^{\infty}dx\int_0^{\infty}dy\,(x+y)K_1^2(x+y)=\frac{3\pi^2}{32}.
\end{equation}

The temperature dependence in the considered region of temperatures
as well as the dependence on the applied voltage is given by Eq. (\ref{nobdwithscr1}).
Within a comparatively big temperature interval the drag is a descending function of temperature.
At smaller temperatures it reaches a maximum.
At small applied voltages we have
\begin{equation}
\label{nobdwithscr02} J_{\rm drag}=J_0
\frac{eV}{2k_BT}\frac{1}{\cosh^4{[p_n^2/4mk_BT]} },
\end{equation}
a linear dependence on $V$, for bigger voltages the drag current
saturates at
\begin{equation}
\label{nobdwithscr2} J_{\rm drag}=2J_0
\frac{1}{\cosh^2{[p_n^2/4mk_BT]}}.
\end{equation}
We wish to emphasize that the screening has nothing to do with the
temperature dependence. The small factor $(z_0/d)^4$ indicates that the screening can be important as it can explain the
magnitude of the effect (without regard of the screening the
theory would have given too large values of the drag current).

\begin{figure}[htb]
\begin{center}
\includegraphics[width=12cm]{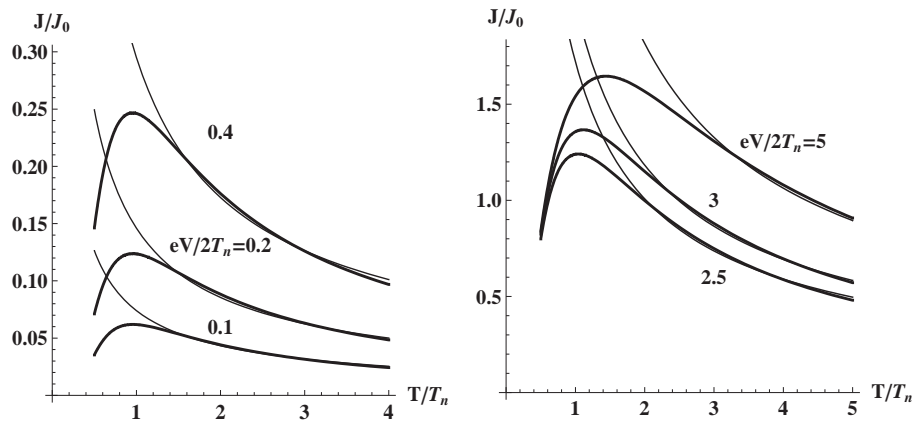}\\
\end{center}
\caption{Temperature dependence of drag current. The thin lines,
corresponding to the $T^{-0.77}$ dependence are plotted for
comparison. \label{pic:dep_on_t_gen}}
\end{figure}

Thus we have come to conclusion that the experimentally observed
temperature dependence can be understood within the Fermi liquid
approach. The temperature dependence is shown in Fig.
\ref{pic:dep_on_t_gen} (for a linear case $eV\,\ll\,T$ on the left of the figure and for large applied voltages $eV\,\gg\,T$ on the right) where $T_n=p_n^2/2m$.
The thin lines correspond to $T^{-0.77}$ law and are given for the
comparison. It is clearly seen that our curves can be also
approximated by the $T^{-0.77}$ dependence, although the authors
of Refs.~\cite{DVRPRM} and~\cite{DZRKVN} regard this dependence as
evidence of Tomonaga-Luttinger liquid behaviour of the quantum
wires. Indeed, they argued that the increase of the drag with
decreasing temperature in a characteristic power-law fashion is in
sharp contrast with the prediction of Fermi liquid theories and,
therefore, may serve as a signature of the TL behaviour.

The Fermi liquid result (see Fig. 1) can be visualized as follows. At very low temperatures there is Fermi degeneracy and therefore the drag
current as a function of temperature goes up. At higher
temperatures the degeneracy is lifted while the average electron
energy increases with temperature. This results in decrease of 
the drag current.

For large values of $n$ and $l$ one can be sure of applicability
of the Fermi liquid approach. In our opinion, it would be of great importance to investigate on experiment and theory physical conditions (including the reservoir influence), that would bring
about transition from the Fermi to Luttinger liquid behavior for
small values of $n$ and $l$. This problem seems to be not simple
since such an investigation should take into account the influence of number of 1D bands in the quantum wire, the vicinity of reservoirs, the electron-phonon interaction, and, of course, the role of temperature.

We would like to point out some outcomes of our theory. First, the interwire
influence can be of importance for the scaled down
devices. According to our theory, to minimize an undesired
influence of this sort one should avoid the alignment of 1D bands.
Second, we note, that as the effect has a maximum as a function of the temperature,
this fact also provides some degree of freedom to change such
influence. On the other hand, the effect can be used as probe in
spectral analysis of nanostructures since it is very
sensitive to the alignment of 1D bands.
And last, the effect can be important for direct investigation of
Coulomb scattering in nanostructures.

\end{document}